\pdfoutput=1

\documentclass[11pt,twoside,a4paper,cmspaper,final,collab]{cms-tdr}
\def\svnVersion{417168}\def\cmsCernNoTag{CERN-EP/2016-277}\def\cmsCernDate{\today}\def\cmsMessage{Published in Physics Letters B as \href{http://dx.doi.org/10.1016/j.physletb.2017.02.012}{\doi{10.1016/j.physletb.2017.02.012.}}}

\begin{document}\cmsNoteHeader{EXO-16-032}

\hyphenation{had-ron-i-za-tion}
\hyphenation{cal-or-i-me-ter}
\hyphenation{de-vices}
\RCS$HeadURL: svn+ssh://alverson@svn.cern.ch/reps/tdr2/papers/EXO-16-032/trunk/EXO-16-032.tex $
\RCS$Id: EXO-16-032.tex 409401 2017-06-09 19:44:52Z rharris $
\newlength\cmsFigWidth
\ifthenelse{\boolean{cms@external}}{\setlength\cmsFigWidth{0.85\columnwidth}}{\setlength\cmsFigWidth{0.4\textwidth}}
\ifthenelse{\boolean{cms@external}}{\providecommand{\cmsLeft}{top\xspace}}{\providecommand{\cmsLeft}{left\xspace}}
\ifthenelse{\boolean{cms@external}}{\providecommand{\cmsRight}{bottom\xspace}}{\providecommand{\cmsRight}{right\xspace}}
\providecommand{\CL}{CL\xspace}
\ifthenelse{\boolean{cms@external}}{\providecommand{\CLp}{C.L}}{\providecommand{\CLp}{CL}}
\ifthenelse{\boolean{cms@external}}{\providecommand{\NA}{\ensuremath{\cdots}\xspace}}{\providecommand{\NA}{\text{---}\xspace}}

\newcommand{\mjj}{\ensuremath{m_{\mathrm{jj}}}\xspace}
\newcommand{\detajj}{\ensuremath{\abs{\Delta\eta_{\mathrm{jj}}}}\xspace}
\newcommand{\Qstar}{\ensuremath{\PQq^*}\xspace}
\newcommand{\RunLumi}{12.9\fbinv}
\newcommand{\lumiUncert}{6.2\%\xspace}
\newcommand{\jecUncert}{2\%\xspace}
\newcommand{\RECOminMjjCut }{1.06\TeV}
\newcommand{\CALOminMjjCut }{0.45\TeV}
\newcommand{\highestMass }{7.7\TeV}
\newcommand{\minMassLow}{0.6\TeV}
\newcommand{\mDM}{\ensuremath{m_{\text{DM}}}\xspace}
\newcommand{\gDM}{\ensuremath{g_{\text{DM}}}\xspace}
\newcommand{\gq}{\ensuremath{g_{\mathrm{q}}}\xspace}
\newcommand{\mMed}{\ensuremath{M_{\text{Med}}}\xspace}
\providecommand{\HT}{\ensuremath{H_\mathrm{T}}\xspace}

\cmsNoteHeader{EXO-16-032}
\title{Search for dijet resonances in proton-proton collisions
at $\sqrt{s}=13$\TeV and constraints on dark matter and other models}

\date{\today}

\abstract{

A search is presented for narrow resonances decaying to dijet final states in proton-proton collisions
at $\sqrt{s}=13$\TeV using data corresponding to an integrated luminosity of 12.9\fbinv.
The dijet mass spectrum is well described by a smooth parameterization and no
significant evidence for the production of new particles is observed.
Upper limits at 95\% confidence level are reported on the production cross
section for narrow resonances with masses above 0.6\TeV.
In the context of specific models, the limits exclude string resonances with masses below
7.4\TeV, scalar diquarks below 6.9\TeV, axigluons and colorons below 5.5\TeV, excited quarks below 5.4\TeV, color-octet
scalars below 3.0\TeV, $\PWpr$ bosons below 2.7\TeV, $\PZpr$ bosons below 2.1\TeV and between
2.3 and 2.6\TeV,
and RS gravitons below 1.9\TeV. These extend previous limits in the dijet channel.
Vector and axial-vector mediators in a simplified model of interactions between quarks and dark matter are excluded below 2.0\TeV.
The first limits in the dijet channel on dark matter mediators are presented as functions of dark matter mass and are compared to
the exclusions of dark matter in direct detection experiments.
}

\hypersetup{%
pdfauthor={CMS Collaboration},%
pdftitle={Search for dijet resonances in proton-proton collisions
at sqrt(s) = 13 TeV and constraints on dark matter and other models},%
pdfsubject={CMS},%
pdfkeywords={CMS, physics, search, exotica, dijet, resonance}}

\maketitle

\section{Introduction}

The dijet mass (\mjj) spectrum in proton-proton ($\Pp\Pp$) collisions arising from the production of partons at high transverse momentum (\pt) 
is predicted by quantum chromodynamics (QCD) to fall smoothly with increasing dijet mass.
Many models of physics beyond the standard model (SM) require new particles that couple to quarks (\PQq) and gluons (\Pg) and can
be observed as resonances in the dijet mass spectrum.
One example is a model in which dark matter (DM) particles couple to quarks
through a DM mediator.  This mediator can decay to either a pair of DM particles or a pair of jets
and therefore can be observed as a dijet resonance~\cite{Chala:2015ama}.
Here, we report a search for narrow dijet resonances, which are those with natural widths that are small compared
to the experimental mass resolution.

This letter presents the results of two searches for dijet resonances, using data collected in 2016 with the CMS detector at the CERN
LHC in $\Pp\Pp$ collisions
at $\sqrt{s}=13$\TeV, corresponding to an integrated luminosity of \RunLumi.
The first is a \textit{high-mass} search for resonances with mass above 1.6\TeV using dijet events that are reconstructed offline.
Similar high-mass searches were published by CMS and ATLAS at $\sqrt{s}=13$\TeV~\cite{Khachatryan:2015dcf,ATLAS:2015nsi},
8\TeV~\cite{Khachatryan:2015sja,Aad:2014aqa,Chatrchyan:2013qhXX}, and
7\TeV~\cite{CMS:2012yf,Aad201237,ATLAS:2012pu,Chatrchyan2011123,Aad:2011aj,Khachatryan:2010jd,ATLAS2010} using
strategies reviewed in Ref.~\cite{Harris:2011bh}. The most recently published high-mass searches used data collected in 2015 corresponding to an
integrated luminosity of 2.4\fbinv by CMS~\cite{Khachatryan:2015dcf} and 3.6\fbinv by ATLAS~\cite{ATLAS:2015nsi}.
The second is a \textit{low-mass} search for resonances with mass between 0.6 and 1.6\TeV
using dijet events that are reconstructed, selected, and recorded in a compact form by the high-level trigger (HLT) in
a technique called \textit{data scouting}~\cite{CMS:2012ooa}. Data scouting was previously used for a similar low-mass search published
by CMS at $\sqrt{s}=8$\TeV~\cite{Khachatryan:2016ecr}.

We present model-independent results and, in addition, consider the following benchmark models of
$s$-channel dijet resonances: string resonances~\cite{Anchordoqui:2008di,Cullen:2000ef}, scalar diquarks~\cite{ref_diquark},  axigluons~\cite{ref_axi,Chivukula:2013xla},
colorons~\cite{ref_coloron,Chivukula:2013xla}, excited quarks
(\Qstar)~\cite{ref_qstar,Baur:1989kv}, color-octet scalars~\cite{Han:2010rf},
new gauge bosons ($\PWpr$ and $\PZpr$) with SM-like or leptophobic couplings~\cite{ref_gauge},
DM mediators~\cite{Boveia:2016mrp,Abdallah:2015ter}, and Randall--Sundrum (RS) gravitons
(G)~\cite{ref_rsg}.
In the color-octet scalar model the squared anomalous coupling used is $k_\mathrm{s}^2=1/2$~\cite{Chivukula:2014pma},
yielding a width and a cross section that is half the value used in the previous CMS search~\cite{Khachatryan:2015dcf}.
Following the recommendations of Ref.~\cite{Boveia:2016mrp} the DM mediator in a simplified model~\cite{Abdallah:2015ter} is assumed to be a spin-1 particle and to decay only to $\PQq\PAQq$ and pairs of DM particles, with unknown mass \mDM, and with
a universal quark coupling $\gq = 0.25$ and a DM coupling $\gDM=1.0$.
Otherwise, the specific choices of
parameters for the benchmark models are the same as those that were used in previous CMS searches, and can
be found in Ref.~\cite{CMS:2012yf}.

\section{Jet reconstruction and event selection}

The CMS detector and its coordinate system, including the azimuthal angle $\phi$ and the pseudorapidity $\eta$,
are described in detail in Ref.~\cite{refCMS}.
The central feature of the CMS apparatus is a superconducting
solenoid of 6 m internal diameter providing an axial field of 3.8 T.
Within the field volume
are located the silicon pixel and strip tracker ($\abs{\eta}<2.4$) and the barrel and
endcap calorimeters ($\abs{\eta}<3$), which consist of a lead tungstate crystal electromagnetic
calorimeter, and a brass and scintillator hadron calorimeter.
An iron and quartz-fiber hadron calorimeter is located in the forward region ($3<\abs{\eta}<5$),
outside the field volume.
For triggering purposes and to facilitate jet reconstruction, the calorimeter
cells are grouped into towers projecting radially outward from the
center of the detector.

A particle-flow (PF) event algorithm reconstructs and identifies each individual particle with an optimized
combination of information from the various elements of the CMS detector~\cite{PFT-09-001-PAS,PFT-10-001-PAS}.
Particles are classified as muons, electrons, photons, and either charged or neutral hadrons. Jets are reconstructed
either using particle flow, giving \textit{PF-jets}, or from energy deposits in the calorimeters,
giving \textit{Calo-jets}. PF-jets reconstructed offline are used in the high-mass
search, and Calo-jets reconstructed by the HLT are used in the low-mass search.
To reconstruct both types of jets, we use the anti-\kt algorithm~\cite{antikt,Cacciari:2005hq} with a distance
parameter of 0.4, as implemented in the \textsc{FastJet} package~\cite{jetarea_fastjet}.
For the high-mass search, at least one reconstructed vertex is required.
The primary vertex is defined as the vertex with the highest sum of $\pt^2$ of the associated tracks.
For PF-jets, charged PF candidates not originating from the primary vertex
are removed prior to the jet finding. For both types of jets, an event-by-event correction based on jet
area~\cite{jetarea_fastjet_pu,Khachatryan:2016kdb}
is applied to the jet energy to remove the estimated contribution from additional collisions in
the same or adjacent bunch crossings (pileup).

Events are selected using a two-tier trigger system. Events satisfying
loose jet requirements at the first level (L1) are examined by the HLT. The HLT
uses \HT, the scalar sum of the jet \pt from all jets in the event
with $\abs{\eta}<3$ that satisfy a jet \pt requirement, to select events. For the high-mass search, PF-jets with $\pt>30$\GeV are used to compute
\HT,
and events are accepted by the HLT if they satisfy the requirement $\HT>800$\GeV.
We then select events with $\mjj>\RECOminMjjCut$ for which the combined L1 trigger
and HLT are found to be fully efficient.
For the low-mass search, when an event passes the HLT, the Calo-jets
reconstructed at the HLT are saved, along with the event energy density and missing transverse momentum
reconstructed from the calorimeter.
The shorter time for event reconstruction of calorimeter quantities and the reduced event size recorded for these events allow a reduced
\HT threshold compared
to the high-mass search. For the low-mass search, Calo-jets with $\pt>40$\GeV are used to compute \HT, the threshold is
$\HT>250$\GeV, and we select events with $\mjj>\CALOminMjjCut$ for which the trigger
is fully efficient.

The jet momenta and energies are corrected using calibration constants
obtained from simulation, test beam results, and pp collision
data at $\sqrt{s}=13$\TeV. The methods described in Ref.~\cite{Khachatryan:2016kdb} are used and all \textit{in-situ}
calibrations are obtained from the current data. All jets are
required to have $\pt>30$\GeV and $\abs{\eta}<2.5$.  The two jets with largest
\pt are defined as the leading jets.
Jet identification (ID) criteria are
applied to remove spurious jets associated with calorimeter
noise. The jet ID for PF-jets is described in Ref.~\cite{JME-10-003-PAS}.
The jet ID for Calo-jets requires that the jet be detected by both the electromagnetic and hadronic
calorimeters with the fraction of jet energy deposited within
the electromagnetic calorimeter between 5 and 95\% of the
total jet energy. An event is rejected if either of the two leading jets fails the jet ID criteria.

Spatially close jets are combined into ``wide jets'' and
used to determine the dijet mass, as in the previous CMS
searches~\cite{Chatrchyan2011123,CMS:2012yf,Chatrchyan:2013qhXX,Khachatryan:2015sja}.  The wide-jet algorithm, designed
for dijet resonance event reconstruction, reduces the analysis sensitivity to gluon radiation from the
final-state partons.  The two leading jets are used as seeds and the four-vectors of all other jets, if within $\Delta R=\sqrt{\smash[b]{(\Delta\eta)^2 +
  (\Delta\phi)^2}}<1.1$, are added to the
nearest leading jet to obtain two wide jets, which then form
the dijet system. The background from $t$-channel dijet events peaks at large values of $\detajj$ and is
suppressed by requiring the pseudorapidity separation of the two wide
jets to satisfy $\detajj<1.3$.
The above requirements maximize the search sensitivity for
isotropic decays of dijet resonances in the presence of QCD dijet background.
For the low-mass search, after wide jet reconstruction and event selection, we use a correction derived from a smaller sample
of dijet data to calibrate the wide jets reconstructed from Calo-jets at HLT. With this correction, based on a 
dijet balance tag-and-probe method similar to that discussed in Ref.~\cite{Khachatryan:2016kdb}, the wide jets from Calo-jets have 
the same response as those reconstructed from PF-jets.

\section{Dijet mass spectrum and fit}

\begin{figure}[htbp]
  \centering
    \includegraphics[width=0.48\textwidth]{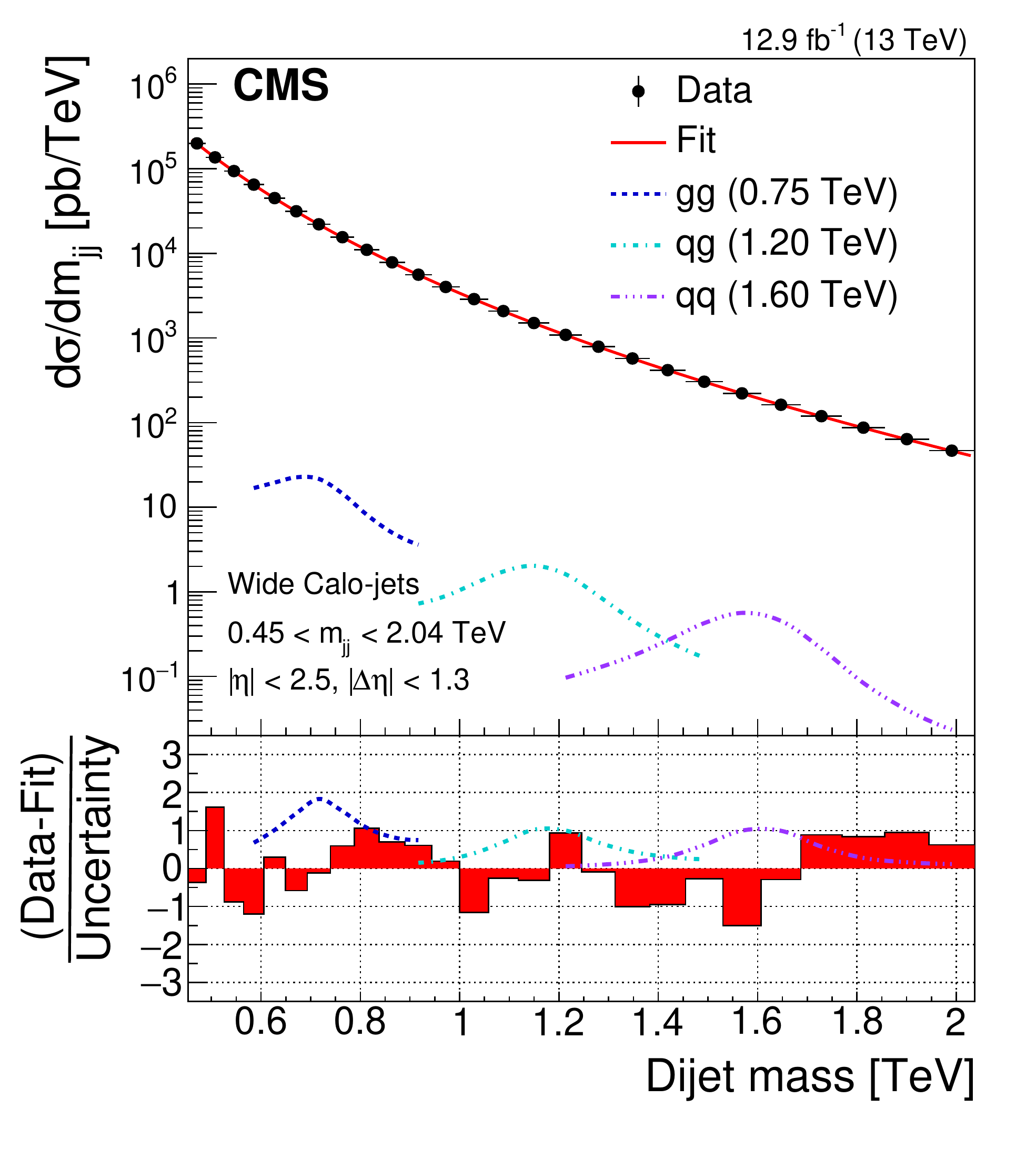}
    \includegraphics[width=0.48\textwidth]{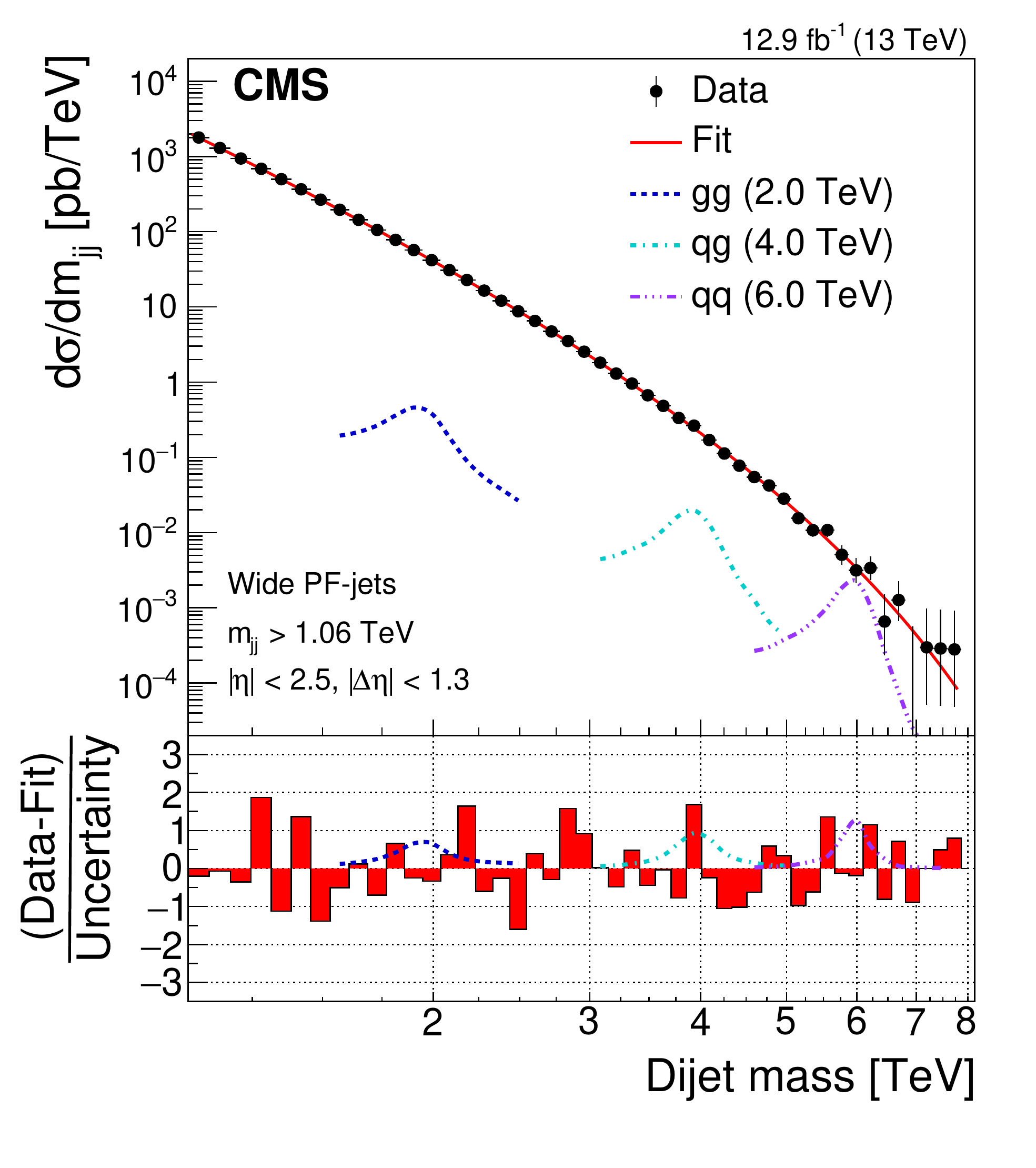}
   \caption{
    Dijet mass spectra (points) compared to a fitted
  parameterization of the background (solid curve) for the low-mass search (\cmsLeft) and
  the high-mass search (\cmsRight).
  The lower panel in each plot shows the difference between the data and the
  fitted parametrization, divided by the statistical uncertainty of the data. Predicted
  signals from narrow gluon-gluon, quark-gluon, and quark-quark resonances are shown with cross sections
  equal to the observed upper limits at 95\% CL.}
    \label{figDataAndFit}
\end{figure}

Figure~\ref{figDataAndFit} shows
the dijet mass spectra, defined as the observed number of events in each bin divided by the
integrated luminosity and the bin width, with predefined bins of width corresponding to the
dijet mass resolution~\cite{Khachatryan:2010jd}.
The highest mass event has a dijet mass of \highestMass.
The dijet mass spectra for both the high- and low-mass searches are fit with the
following parameterization:
\begin{equation}
\frac{{\rd}\sigma}{{\rd}\mjj} =
\frac{P_{0} (1 - x)^{P_{1}}}{x^{P_{2} + P_{3} \ln{(x)}}}\ ,
\label{eqBackgroundParam}
\end{equation}
where $x=\mjj/\sqrt{s}$ and $P_0$, $P_1$, $P_2$, and $P_3$ are four free parameters.
The functional form in Eq.~(\ref{eqBackgroundParam}) was also used in previous
searches \cite{Khachatryan:2016ecr,Khachatryan:2015dcf,Khachatryan:2010jd,Chatrchyan2011123,CMS:2012yf,Chatrchyan:2013qhXX,Khachatryan:2015sja,
ATLAS:2015nsi,ATLAS2010,Aad:2011aj,Aad201237,ATLAS:2012pu,Aad:2014aqa,refCDFrun2} to describe the data.
In Fig.~\ref{figDataAndFit} we show the result of binned maximum likelihood fits, performed independently, which yields the following chi-squared per number
of degrees of freedom: $\chi^2/\mathrm{NDF}=33.3/42$ for the high-mass search and $\chi^2/\mathrm{NDF}=17.3/22$ for the low-mass search.
The dijet mass spectra are well modeled by the background fits.
In the lower panels of Fig.~\ref{figDataAndFit}, in the region of dijet mass between 1.1 and 2.0\TeV,
the bin-by-bin differences between the data and the background fit are not identical in the two searches
because fluctuations in reconstructed dijet mass for Calo-jets and PF-jets are not completely correlated.

We search for narrow resonances in the dijet mass spectrum.
Figure~\ref{figDataAndFit} shows examples of dijet mass distributions
for signal events generated with the \PYTHIA8.205~\cite{Sjostrand:2007gs} program with the
CUETP8M1 tune~\cite{Khachatryan:2015pea,Skands:2014pea} and including a \GEANTfour-based \cite{refGEANT} simulation of the CMS detector.
The predicted mass distributions have Gaussian cores from jet energy resolution,
and tails towards lower mass values primarily from QCD radiation. The
contribution of the low mass tail to the lineshape depends on the
parton content of the resonance ($\PQq\PQq$, $\PQq\Pg$, or $\Pg\Pg$).  Resonances
containing gluons, which emit more QCD radiation than
quarks, are wider and have a more pronounced tail.
The signal distributions shown in Fig.~\ref{figDataAndFit} are for $\PQq\PQq$, $\PQq\Pg$, and $\Pg\Pg$
resonances with signal cross sections corresponding to the limits at 95\% confidence level (CL) obtained by this analysis, as described below.
There is no evidence for a narrow resonance in the data.
The most significant excess of the data relative to the background fit
comes from the five consecutive bins between 0.74 and 1.00\TeV in the low mass search shown in Fig.~\ref{figDataAndFit}.
Fitting these data to $\PQq\PQq$, $\PQq\Pg$, and $\Pg\Pg$ resonances with a mass of 0.85\TeV
yields local significances of 2.2, 2.5 and 2.6 standard deviations including systematic uncertainties, respectively.

\section{Limits on dijet resonances}

We use the dijet mass spectrum from wide jets, the background
parameterization, and the dijet resonance shapes to set
limits on the production of new particles decaying to the parton pairs $\PQq\PQq$ (or $\PQq\PAQq$), $\Pq\Pg$, and $\Pg\Pg$. A separate limit is
determined for each final state ($\PQq\PQq$, $\Pq\Pg$, and $\Pg\Pg$) because of the dependence of the
dijet resonance shape on the types of the two final-state partons.

The dominant sources of systematic uncertainty are the jet energy scale and resolution,
integrated luminosity, and the estimation of background. The uncertainty in the jet energy scale in both the
low-mass and the high-mass search is \jecUncert and is
determined from $\sqrt{s}=13$\TeV data using the methods described in Ref.~\cite{Khachatryan:2016kdb}.
This uncertainty is propagated to the limits by shifting the dijet mass shape for signal by $\pm$\jecUncert.
The uncertainty in the jet energy
resolution translates into an uncertainty of 10\% in the resolution of the dijet
mass~\cite{Khachatryan:2016kdb}, and is propagated to the limits by observing the effect of increasing and decreasing by 10\% the reconstructed
width of the dijet mass shape for signal.
The uncertainty in the
integrated luminosity is \lumiUncert, and is propagated to the normalization of
the signal.
Changes in the values of the parameters describing the background introduce a change in the signal strength,
which is accounted for as a systematic uncertainty as discussed in the next paragraph.

The modified frequentist method~\cite{Junk1999,bib-cls} is
utilized to set upper limits on signal cross sections, following the prescription
described in Refs.~\cite{ATLAS:1379837,Cowan:2010js}.  We use a multi-bin counting experiment likelihood, which is
a product of Poisson distributions corresponding to different bins.
We evaluate the likelihood independently at each value of resonance pole mass from 0.6 to 1.6\TeV in 50-\GeVns
steps in the low-mass search, and from 1.6 to 7.5\TeV in 100-\GeVns steps in the high-mass search.
The systematic uncertainties are implemented as nuisance parameters in the likelihood model, with Gaussian constraints for the jet
energy scale and resolution, and log-normal constraints for the integrated luminosity.
The systematic uncertainty in the background is automatically evaluated via profiling, effectively
refitting for the optimal values of the background parameters for each value of resonance cross section.
This procedure gives the same limits as the Bayesian procedure used previously for dijet resonance
searches at CMS~\cite{Khachatryan:2015sja}. For both the Bayesian and modified frequentist statistical procedures
we find that the background systematic uncertainty has the largest effect on the limit. The extent to which the
background uncertainty affects the limit depends significantly on the signal shape and the resonance mass, with the largest
effect occurring for
the $\Pg\Pg$ resonances because they are wider, and the smallest effect for $\PQq\PQq$ resonances.
The effect decreases as the resonance mass increases. For example, considering two signals shown in
Fig.~\ref{figDataAndFit}: for a $\Pg\Pg$ resonance at a mass of 0.75\TeV systematic uncertainties increase the limit by a factor of 3,
and for a $\PQq\PQq$ resonance at a mass of 6\TeV systematic uncertainties increase the limit by only 10\%.
\begin{figure*}[hbtp!]
  \centering
    \includegraphics[width=0.48\textwidth]{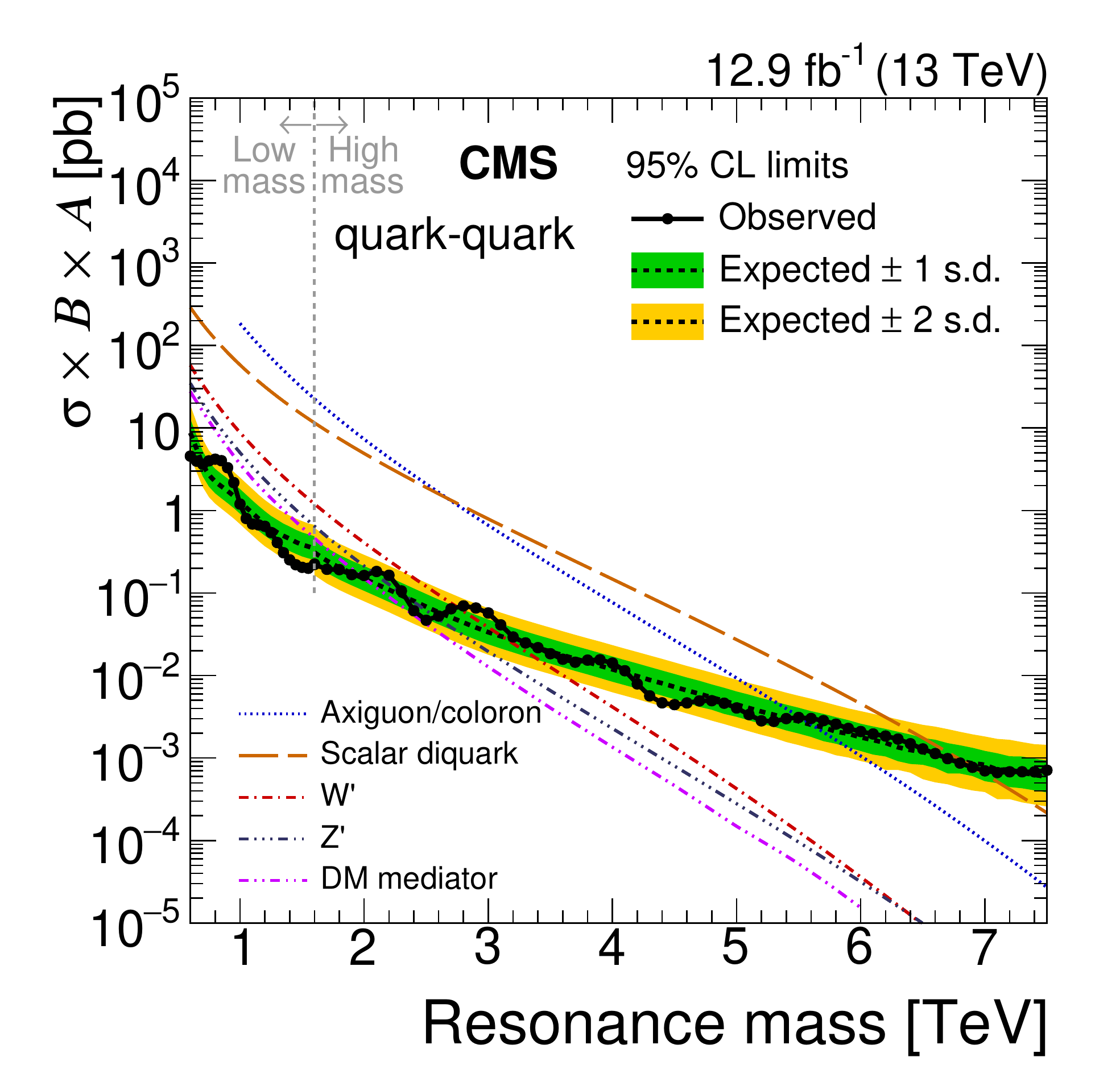}
    \includegraphics[width=0.48\textwidth]{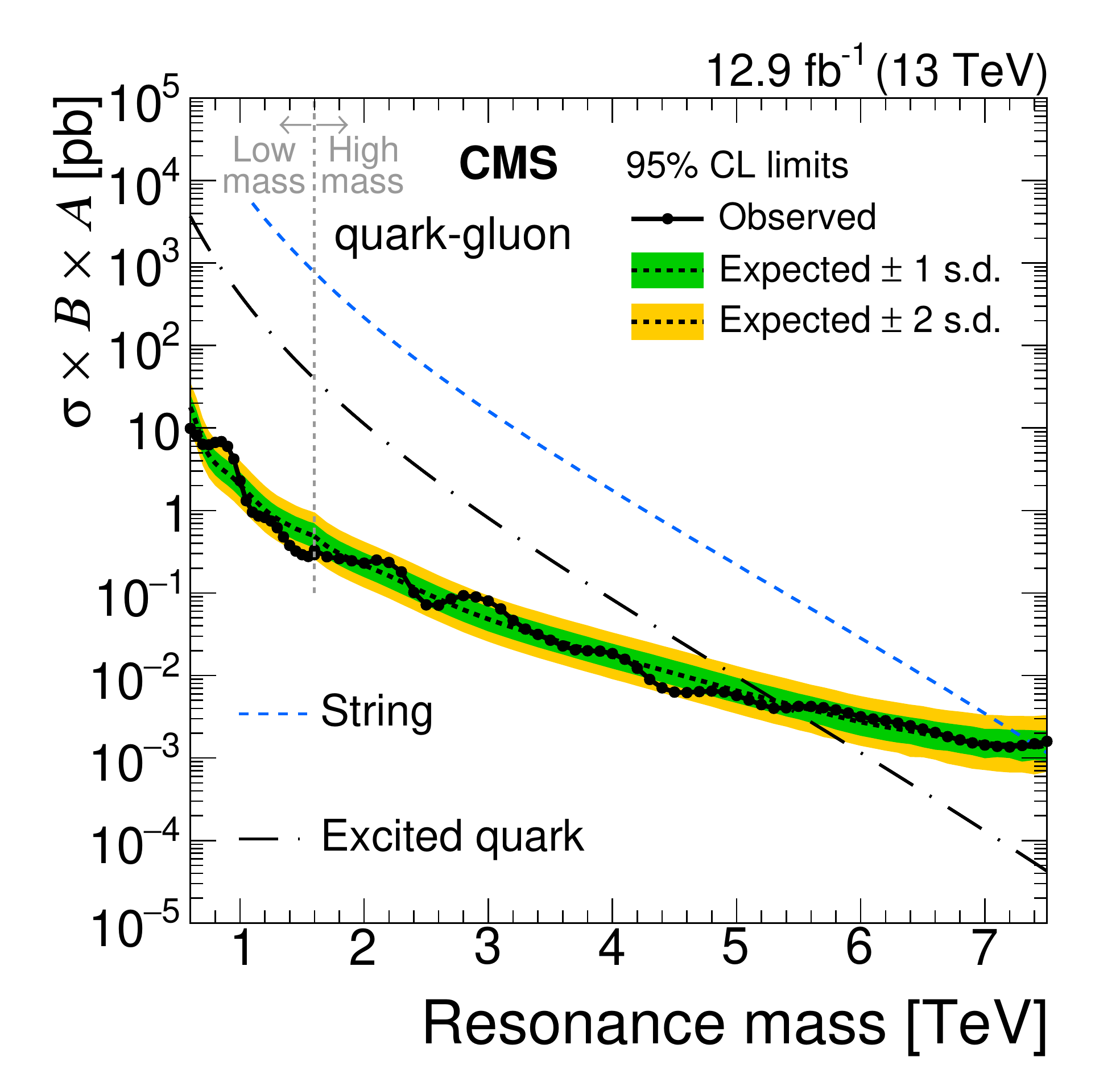}
    \includegraphics[width=0.48\textwidth]{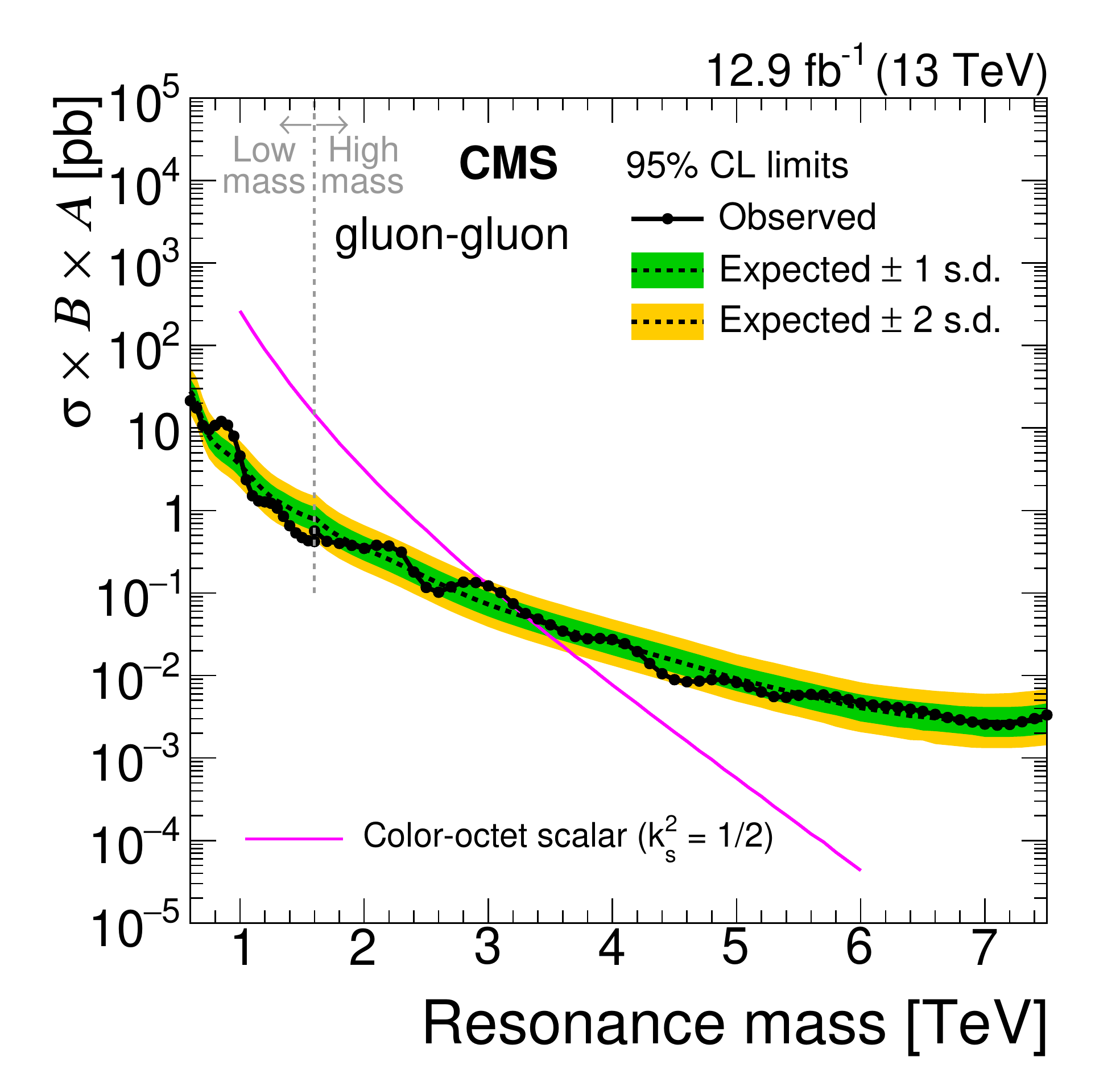}
    \includegraphics[width=0.48\textwidth]{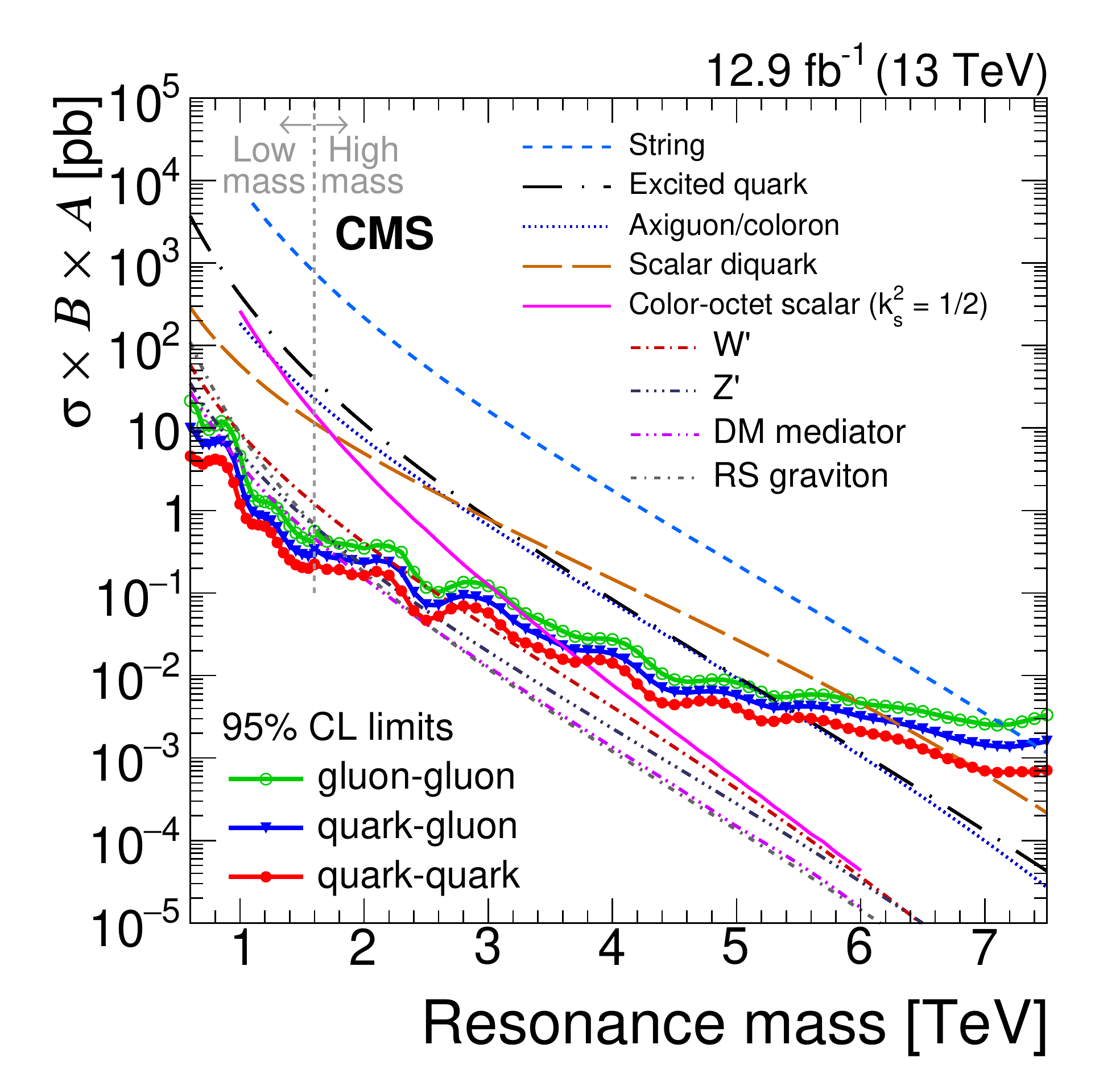}
    \caption{The observed 95\% \CL upper limits on the product of the cross section, branching fraction, and acceptance for
    quark-quark (top left), quark-gluon (top right), and gluon-gluon (bottom left) type dijet resonances.
    The corresponding expected limits (dashed) and their variations
    at the 1 and 2 standard deviation levels (shaded bands) are also shown.
    All observed limits (solid) are compared (bottom right). Limits are compared to predicted cross sections for string resonances~\cite{Anchordoqui:2008di,Cullen:2000ef},
    excited quarks~\cite{ref_qstar,Baur:1989kv}, axigluons~\cite{ref_axi}, colorons~\cite{ref_coloron},  scalar diquarks~\cite{ref_diquark},
    color-octet scalars~\cite{Han:2010rf}, new gauge bosons $\PWpr$ and $\PZpr$ with SM-like couplings~\cite{ref_gauge},
    dark matter mediators for $m_{\mathrm{DM}}=1$\GeV~\cite{Boveia:2016mrp,Abdallah:2015ter}, and RS gravitons~\cite{ref_rsg}.
    }
    \label{figLimitAll}
\end{figure*}

Signal injection tests were performed to investigate the potential bias introduced through the choice of
background parameterization. Pseudo-data generated assuming an alternative
parameterization, ${\rd}\sigma/\rd\mjj=\exp(\ln{(P_0)} + P_1x^{P_2} + P_1(1-x)^{P_3})$, were fit with the nominal parameterization
given in Eq.~(\ref{eqBackgroundParam}). The bias in the extracted signal
was found to be negligible. We tried other functions but did not find any with four or fewer parameters that could fit our data.

Figure~\ref{figLimitAll} shows the model-independent observed
upper limits at 95\% \CL on the product of
the cross section ($\sigma$), the branching fraction ($B$), and the acceptance ($A$) for narrow resonances, with the
kinematic requirements $\detajj<1.3$ and $\abs{\eta}<2.5$. The acceptance of the minimum dijet mass requirement in each search has been evaluated separately for 
$\PQq\PQq$, $\PQq\Pg$, and $\Pg\Pg$ resonances, and has been taken into account by correcting the limits, and therefore does not appear in the acceptance $A$.
The corrections are independent of the spin and coupling of the narrow resonance at the one percent level. 
Figure \ref{figLimitAll} also shows the expected limits on the cross section and their bands of uncertainty.
The difference in the limits for $\PQq\PQq$, $\PQq\Pg$, and $\Pg\Pg$ resonances at the same resonance mass
originates from the difference in their lineshapes.

All upper limits presented can be compared to the parton-level predictions of $\sigma\, B\, A$, without detector simulation,
to determine mass limits on new particles.
The model predictions shown in Fig.~\ref{figLimitAll} are calculated in the narrow-width
approximation~\cite{Harris:2011bh} using the CTEQ6L1~\cite{refCTEQ} PDF at leading order,
with a next-to-leading order correction factor of approximately 1.3 included for the $\PWpr$ and $\PZpr$ models, and approximately 1.2 for
the axigluon/coloron models~\cite{Chivukula:2013xla}.
The branching fraction includes the direct decays of the resonance into the five light quarks and
gluons only, excluding top quarks from the decay, although top quarks are included in the calculation of
the resonance width.
The acceptance is evaluated at the parton level for the resonance decay to two partons. In the case of isotropic
decays, the acceptance is $A\approx 0.6$ and is independent of the resonance mass.
For a given model, new particles are excluded at 95\% \CL in mass regions where the theoretical prediction
lies at or above the observed upper limit for the appropriate final state of Fig.~\ref{figLimitAll}.
For the RS graviton model, the decay fraction is 60\% to quarks and 40\% to gluons, and we obtain mass limits by comparing
the model cross section curve to the weighted average of the limits
in the $\PQq\PQq$ and $\Pg\Pg$ final states.
Mass limits on all benchmark models are summarized in Table~\ref{tab:MassLimit} and are more stringent than the mass limits
in the dijet channel previously published by CMS~\cite{Khachatryan:2015dcf} and ATLAS~\cite{ATLAS:2015nsi}.

\begin{table}[htb]
  \topcaption{ Observed and expected mass limits at 95\% CL. The listed models are excluded between \minMassLow
  and the indicated mass. In addition to the observed mass limits
  listed below, this analysis also excludes a $\PZpr$ in the mass interval between 2.3 and 2.6 \TeV.}
  \centering
  \begin{tabular}{lcll}
    Model & Final                            & \multicolumn{2}{c}{Limit [\TeVns{}]} \\
         &   state                                & Obs.      & Exp.      \\
    \hline
String   & $\PQq\Pg$                        & 7.4   & 7.4\\
Scalar diquark  & $\PQq\PQq$                & 6.9   & 6.8\\
Axigluon/coloron  & $\PQq\PAQq$             & 5.5   & 5.6\\
Excited quark  &  $\PQq\Pg$                 & 5.4   & 5.4\\
Color-octet scalar ($k_s^2=1/2$) & $\Pg\Pg$ & 3.0   & 3.3\\
$\PWpr$ & $\PQq\PAQq$                       & 2.7   & 3.1\\
$\PZpr$ & $\PQq\PAQq$                       & 2.1   & 2.3\\
DM mediator ($m_{\mathrm{DM}}=1$\GeVns{}) & $\PQq\PAQq$ & 2.0   & 2.0 \\
RS graviton  & $\PQq\PAQq$, $\Pg\Pg$        & 1.9   & 1.8\\
 \end{tabular}
\label{tab:MassLimit}
\end{table}

\begin{figure}[hbtp]
  \centering
    \includegraphics[width=0.48\textwidth]{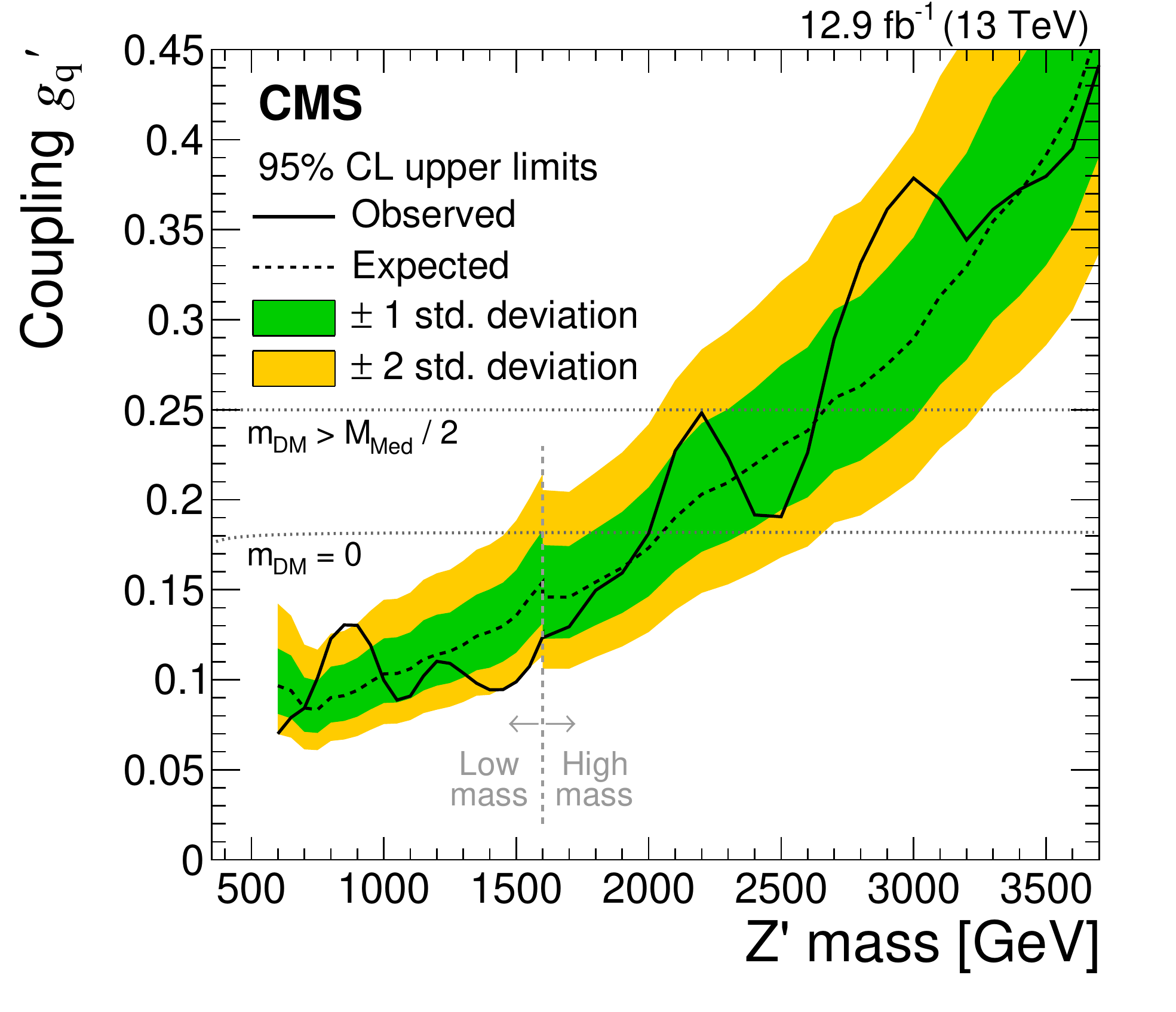}
\caption{  The 95\% \CL upper limits on the universal
quark coupling $\gq^{\prime}$ as a function of resonance mass for a leptophobic $\PZpr$ resonance that only couples to quarks.
The observed limits (solid), expected
limits (dashed) and their variation at the 1 and 2 standard deviation levels (shaded bands) are shown.
Dotted horizontal lines show the coupling strength for which the cross section for dijet production in this model is the same
as for a DM mediator (see text).}
    \label{figCoupling}
\end{figure}

Mass limits on new particles are sensitive to assumptions about their coupling.
Conversely, at a fixed resonance mass, models with smaller couplings are excluded by searches with increased sensitivity.
Figure~\ref{figCoupling} shows our upper limits on the coupling as a function of
mass for a model of a leptophobic $\PZpr$ resonance with a universal quark coupling,
$\gq^{\prime}$~\cite{Boveia:2016mrp}, related to the $\PZpr$ coupling convention of
Ref.~\cite{Dobrescu:2013coa} by $\gq^{\prime}=g_B/6$.

\section{Limits on dark matter}

\begin{figure}[hbtp]
  \centering
    \includegraphics[width=0.48\textwidth]{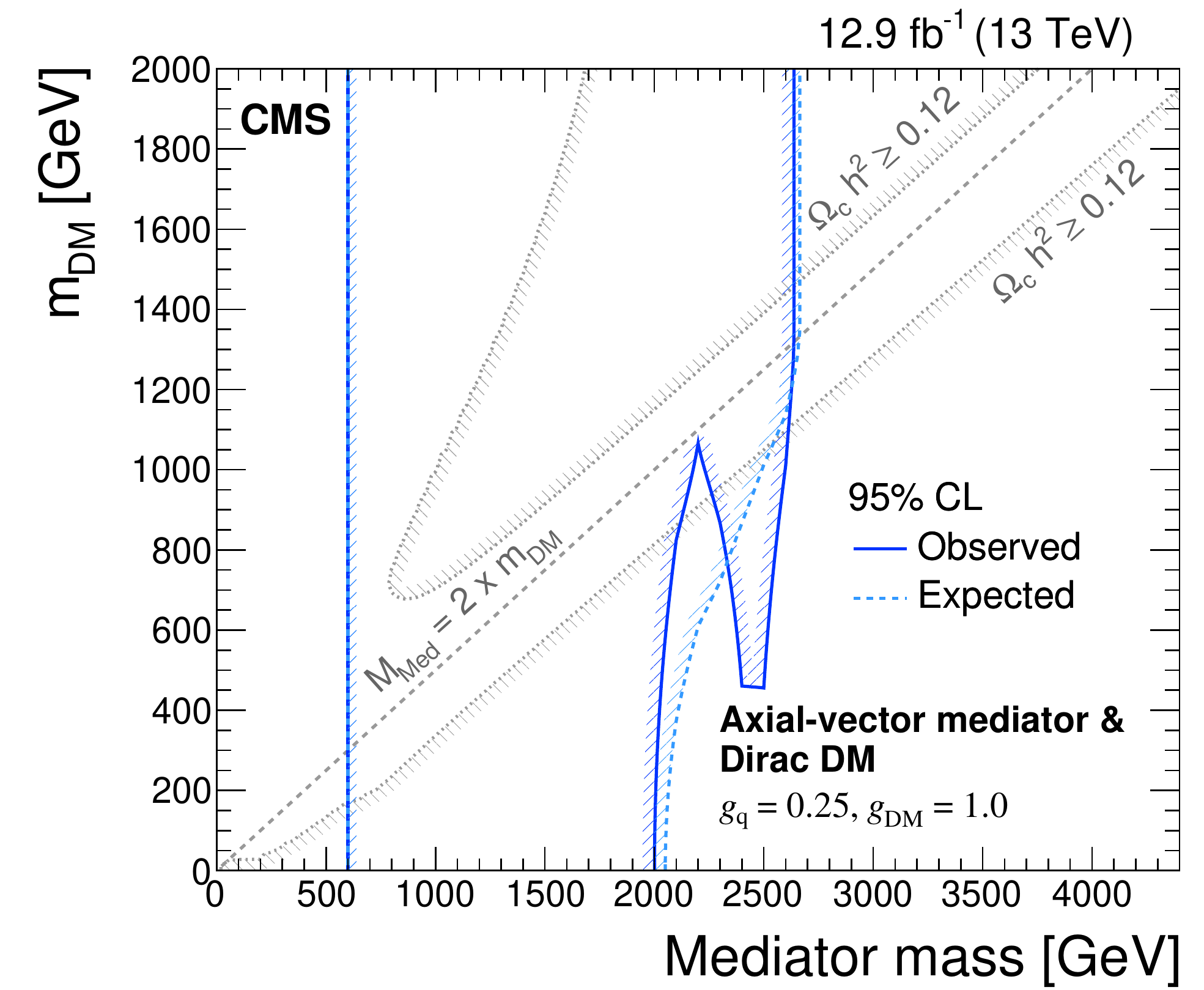}
    \includegraphics[width=0.48\textwidth]{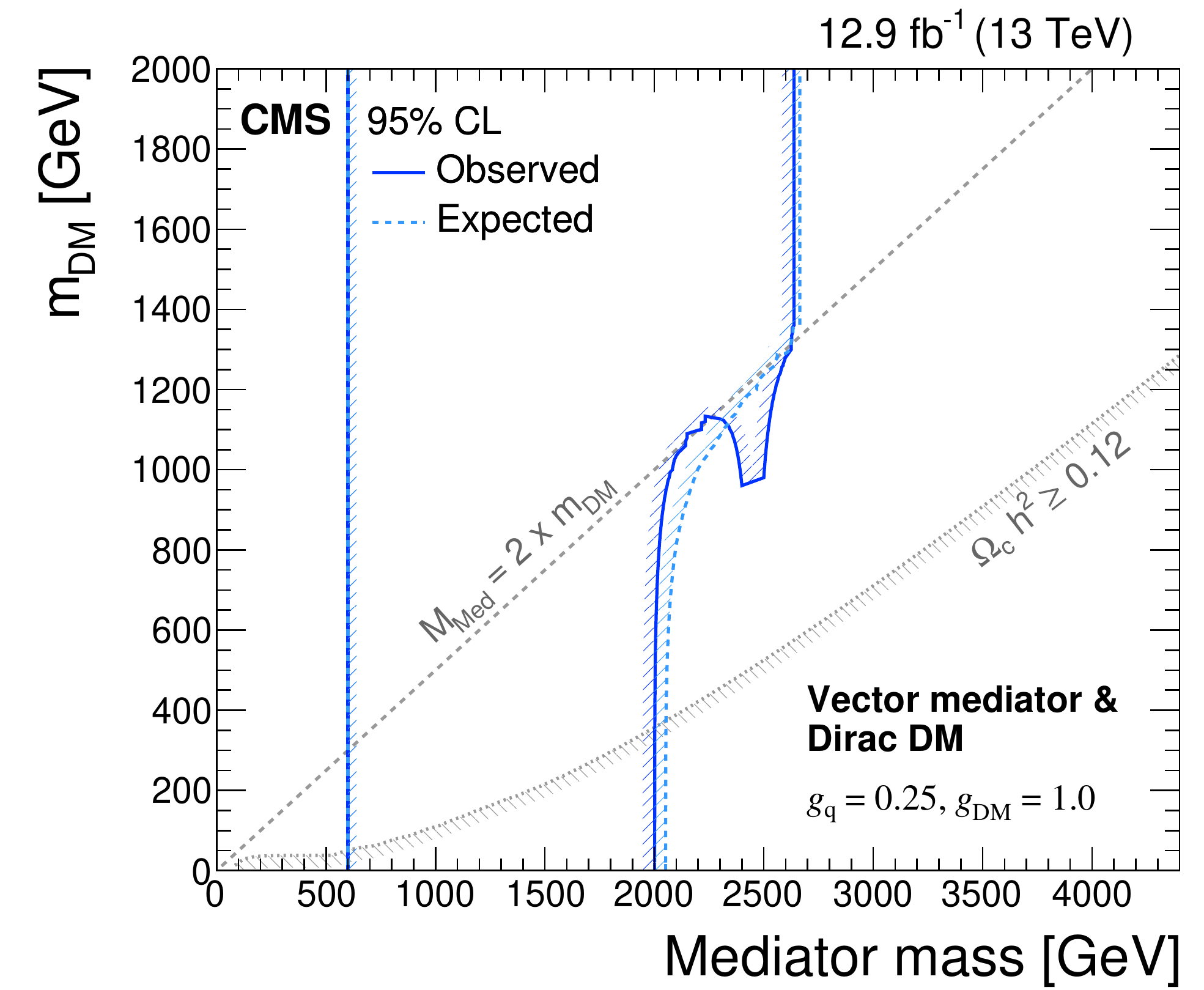}
\caption{  The 95\% CL observed (solid) and expected (dashed) excluded regions in the plane of dark matter mass vs.
mediator mass, for an axial-vector mediator (\cmsLeft) and a vector mediator (\cmsRight), are compared to constraints from the cosmological relic
density of DM (light gray) determined from astrophysical measurements~\cite{Spergel:2006hy,Ade:2013zuv} and \textsc{MadDM} version
2.0.6~\cite{Backovic:2013dpa,Backovic:2015cra} as described in Ref.~\cite{Pree:2016hwc}.
Following the recommendation of the
LHC DM working group~\cite{Boveia:2016mrp, Abdallah:2015ter}, the exclusions are computed for Dirac DM and for a universal quark coupling $\gq = 0.25$ and for a DM
coupling of $\gDM=1.0$. It should also be noted that the excluded region
strongly depends on the chosen coupling and model scenario. Therefore, the excluded regions and relic density contours shown in this plot are
not applicable to other choices of coupling values or models. }
    \label{figDM}
\end{figure}

We use our limits to constrain simplified models of DM, with leptophobic vector and axial-vector
mediators that couple only to quarks and DM particles~\cite{Boveia:2016mrp,Abdallah:2015ter}.
Figure~\ref{figDM} shows the excluded values of mediator mass as a function of \mDM
for both types of mediators.
For \mDM = 1\GeV, indistinguishable from zero, the excluded range of mediator mass (\mMed) is between 0.6 and 2.0\TeV, as
also shown in Fig.~\ref{figLimitAll} and listed in Table~\ref{tab:MassLimit}. An additional excluded range of $0.5<\mMed<0.6$\TeV, not shown, comes from the
low-mass search at $\sqrt{s}=8$\TeV~\cite{Khachatryan:2016ecr}.
In Fig.~\ref{figDM} the expected upper value of excluded \mMed increases with \mDM to as high as 2.65\TeV because
the branching fraction to $\PQq\PAQq$ increases with \mDM.
If $\mDM>\mMed/2$, the mediator cannot decay to DM particles, and the dijet cross section from the mediator models becomes identical to that
in the leptophobic $\PZpr$ model used in Fig.~\ref{figCoupling} with a coupling $\gq^{\prime}=\gq=0.25$. Therefore for these values of \mDM
the limits on the mediator mass in Fig.~\ref{figDM} are identical to the limits on the $\PZpr$ mass at $\gq^{\prime}=0.25$ in Fig.~\ref{figCoupling}.
Similarly, if $\mDM=0$, the limits on the mediator mass in Fig.~\ref{figDM} are identical to the limits on the $\PZpr$ mass at
$\gq^{\prime}=\gq/\sqrt{1+16/(3N_f)}\approx 0.182$ in Fig.~\ref{figCoupling}, where $N_f$ is the effective number of quark flavors contributing
to the width of the resonance.

\begin{figure}[hbtp]
  \centering
    \includegraphics[width=0.48\textwidth]{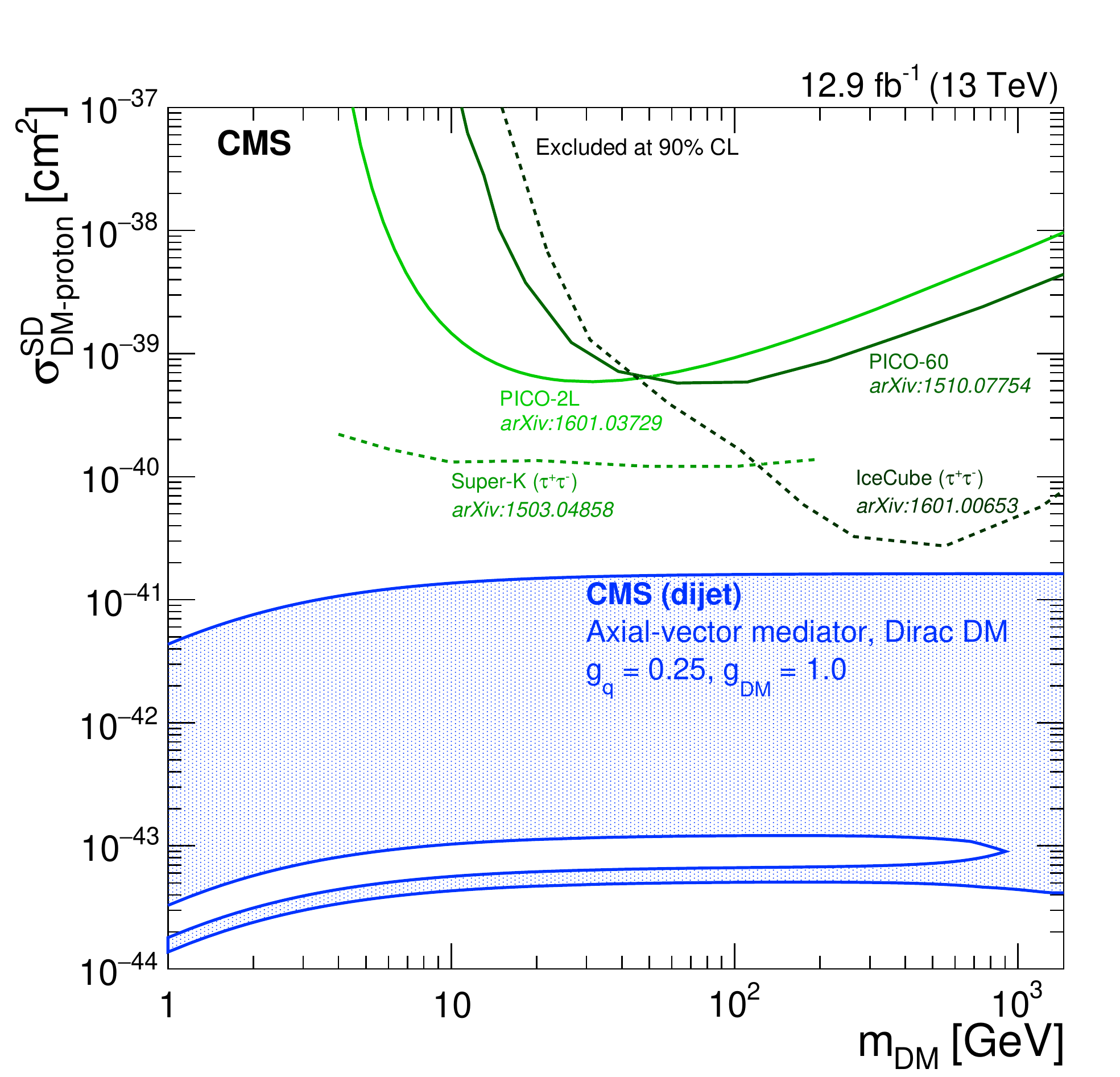}
    \includegraphics[width=0.48\textwidth]{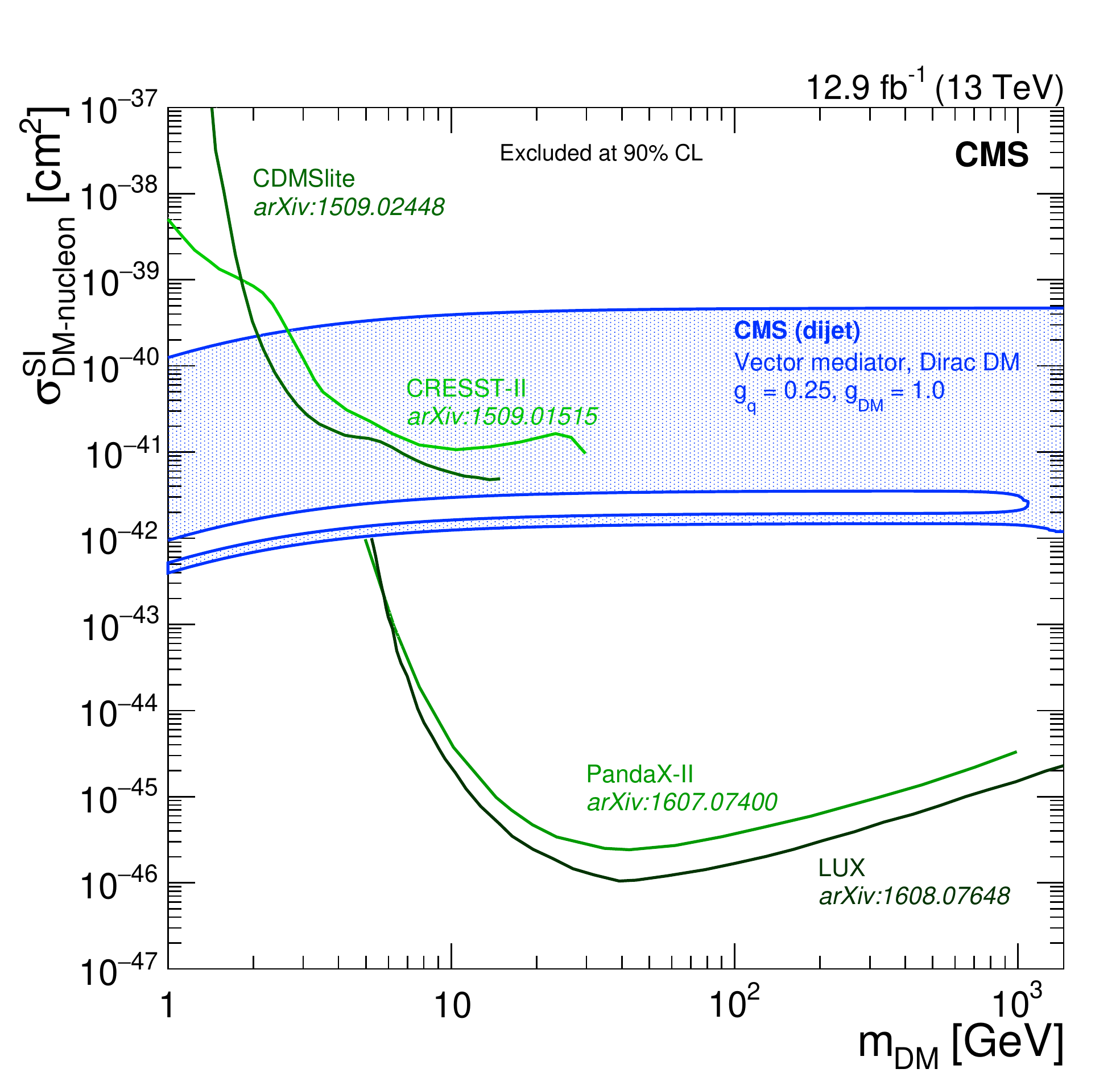}
\caption{Excluded regions at 90\% CL in the plane of dark matter nucleon interaction cross section vs.
dark matter mass. (\cmsLeft) The CMS exclusion of a spin-dependent cross section (shaded) from
an axial-vector mediator decaying to dijets is compared with limits from the PICO experiments~\cite{Amole:2016pye, Amole:2015pla}, IceCube~\cite{Aartsen:2016exj}, and Super-Kamiokande~\cite{Choi:2015ara}.
(\cmsRight)  The CMS exclusion of a spin-independent cross section (shaded) from a vector mediator decaying
to dijets is compared with the LUX 2016~\cite{Akerib:2016vxi}, PandaX-II 2016~\cite{Tan:2016zwf}, CDMSLite 2015~\cite{Agnese:2015nto}, and CRESST-II 2015~\cite{Angloher:2015ewa} limits, which have documented the most constraining results in
the shown mass range.  The CMS exclusions are for Dirac DM and couplings $\gq=0.25$ and $\gDM=1$, for leptophobic axial-vector
and vector mediators, and they strongly depend on these choices and are not applicable to other choices of
coupling values or models. The CMS limits do not include a constraint on the relic density.
 }
\label{figDD}
\end{figure}

As outlined in detail in Ref.~\cite{Boveia:2016mrp} these results can also be
compared with results from direct detection experiments.
The limits in Fig.~\ref{figDM} are first re-calculated at 90\% CL,
and then translated into the plane of the DM mass versus the DM-nucleon interaction
cross section from the predicted relation between the interaction cross section and the mediator mass. An axial-vector mediator leads
to a spin-dependent cross section, $\sigma^\mathrm{SD}$, and a vector mediator leads to a spin-independent cross
section, $\sigma^\mathrm{SI}$. Figure~\ref{figDD} shows the comparison of these results with dark matter searches by direct detection~\cite{Amole:2016pye, Amole:2015pla, Aartsen:2016exj, Choi:2015ara,Akerib:2016vxi,Tan:2016zwf, Agnese:2015nto, Angloher:2015ewa}.
The gap in the CMS excluded region in Fig.~\ref{figDD} corresponds to a structure with a statistical significance of one standard deviation
seen at a mass of 2.2\TeV in Figs.~\ref{figDataAndFit}-\ref{figDM}.
For our benchmark model the present search excludes a significantly smaller $\sigma^\mathrm{SD}$ than the direct detection experiments, and a competitive region
of $\sigma^\mathrm{SI}$. We note that the absolute exclusion of this search, as well as its relative importance with respect to other dark matter searches,
strongly depends on the chosen coupling and model scenario. Nevertheless, this benchmark model, a vector or an axial-vector mediator with
a universal quark coupling $\gq = 0.25$ and a DM coupling of $\gDM=1.0$, illustrates that dijet searches can place significant bounds
on relevant DM models and thus are important ingredients in the search for DM.

\section{Summary}

Two searches for narrow resonances decaying into a pair of jets have been performed using
proton-proton collisions at $\sqrt{s}=13$\TeV corresponding to an integrated luminosity of 12.9\fbinv: a low-mass search
based on calorimeter jets, reconstructed by the high level trigger and recorded in compact form (data scouting),
and a high-mass search based on particle-flow jets. The dijet mass spectra are observed to be smoothly falling
distributions. In the analyzed data samples, there is no evidence for resonant particle production.
Generic upper limits are presented on the product of the cross section, the branching fraction, and the acceptance for narrow quark-quark, quark-gluon, and gluon-gluon
resonances that are applicable to any model of narrow dijet resonance production.
String resonances with masses below 7.4\TeV are excluded at 95\% confidence level, as are scalar diquarks below 6.9\TeV,
axigluons and colorons below 5.5\TeV, excited quarks below 5.4\TeV, color-octet
scalars below 3.0\TeV, $\PWpr$ bosons below 2.7\TeV, $\PZpr$ bosons with SM-like couplings below 2.1\TeV and between 2.3 and 2.6\TeV,
and Randall--Sundrum gravitons below 1.9\TeV. This extends previously published limits in the dijet channel. The first limits
are set on a simplified model of dark matter mediators based on the dijet channel, excluding vector and axial-vector mediators
below 2.0\TeV, and using a universal quark coupling $\gq = 0.25$ and a dark matter coupling $\gDM=1.0$. Limits on the mass of a dark matter mediator are presented as a function of dark matter
mass, and are translated into upper limits on the cross section for dark matter particles scattering on nucleons
that are more sensitive than those of direct detection experiments for spin-dependent cross sections.

\begin{acknowledgments}
We congratulate our colleagues in the CERN accelerator departments for the excellent performance of the LHC and thank the technical and administrative staffs at CERN and at other CMS institutes for their contributions to the success of the CMS effort. In addition, we gratefully acknowledge the computing centers and personnel of the Worldwide LHC Computing Grid for delivering so effectively the computing infrastructure essential to our analyses. Finally, we acknowledge the enduring support for the construction and operation of the LHC and the CMS detector provided by the following funding agencies: BMWFW and FWF (Austria); FNRS and FWO (Belgium); CNPq, CAPES, FAPERJ, and FAPESP (Brazil); MES (Bulgaria); CERN; CAS, MoST, and NSFC (China); COLCIENCIAS (Colombia); MSES and CSF (Croatia); RPF (Cyprus); SENESCYT (Ecuador); MoER, ERC IUT, and ERDF (Estonia); Academy of Finland, MEC, and HIP (Finland); CEA and CNRS/IN2P3 (France); BMBF, DFG, and HGF (Germany); GSRT (Greece); OTKA and NIH (Hungary); DAE and DST (India); IPM (Iran); SFI (Ireland); INFN (Italy); MSIP and NRF (Republic of Korea); LAS (Lithuania); MOE and UM (Malaysia); BUAP, CINVESTAV, CONACYT, LNS, SEP, and UASLP-FAI (Mexico); MBIE (New Zealand); PAEC (Pakistan); MSHE and NSC (Poland); FCT (Portugal); JINR (Dubna); MON, RosAtom, RAS, and RFBR (Russia); MESTD (Serbia); SEIDI and CPAN (Spain); Swiss Funding Agencies (Switzerland); MST (Taipei); ThEPCenter, IPST, STAR, and NSTDA (Thailand); TUBITAK and TAEK (Turkey); NASU and SFFR (Ukraine); STFC (United Kingdom); DOE and NSF (USA).

\hyphenation{Rachada-pisek} Individuals have received support from the Marie-Curie program and the European Research Council and EPLANET (European Union); the Leventis Foundation; the A. P. Sloan Foundation; the Alexander von Humboldt Foundation; the Belgian Federal Science Policy Office; the Fonds pour la Formation \`a la Recherche dans l'Industrie et dans l'Agriculture (FRIA-Belgium); the Agentschap voor Innovatie door Wetenschap en Technologie (IWT-Belgium); the Ministry of Education, Youth and Sports (MEYS) of the Czech Republic; the Council of Science and Industrial Research, India; the HOMING PLUS program of the Foundation for Polish Science, cofinanced from European Union, Regional Development Fund, the Mobility Plus program of the Ministry of Science and Higher Education, the National Science Center (Poland), contracts Harmonia 2014/14/M/ST2/00428, Opus 2013/11/B/ST2/04202, 2014/13/B/ST2/02543 and 2014/15/B/ST2/03998, Sonata-bis 2012/07/E/ST2/01406; the Thalis and Aristeia programs cofinanced by EU-ESF and the Greek NSRF; the National Priorities Research Program by Qatar National Research Fund; the Programa Clar\'in-COFUND del Principado de Asturias; the Rachadapisek Sompot Fund for Postdoctoral Fellowship, Chulalongkorn University and the Chulalongkorn Academic into Its 2nd Century Project Advancement Project (Thailand); and the Welch Foundation, contract C-1845.
\end{acknowledgments}

\bibliography{auto_generated}
\cleardoublepage \appendix\section{The CMS Collaboration \label{app:collab}}\begin{sloppypar}\hyphenpenalty=5000\widowpenalty=500\clubpenalty=5000\textbf{Yerevan Physics Institute,  Yerevan,  Armenia}\\*[0pt]
A.M.~Sirunyan, A.~Tumasyan
\vskip\cmsinstskip
\textbf{Institut f\"{u}r Hochenergiephysik,  Wien,  Austria}\\*[0pt]
W.~Adam, E.~Asilar, T.~Bergauer, J.~Brandstetter, E.~Brondolin, M.~Dragicevic, J.~Er\"{o}, M.~Flechl, M.~Friedl, R.~Fr\"{u}hwirth\cmsAuthorMark{1}, V.M.~Ghete, C.~Hartl, N.~H\"{o}rmann, J.~Hrubec, M.~Jeitler\cmsAuthorMark{1}, A.~K\"{o}nig, I.~Kr\"{a}tschmer, D.~Liko, T.~Matsushita, I.~Mikulec, D.~Rabady, N.~Rad, B.~Rahbaran, H.~Rohringer, J.~Schieck\cmsAuthorMark{1}, J.~Strauss, W.~Waltenberger, C.-E.~Wulz\cmsAuthorMark{1}
% ...
\vskip\cmsinstskip
% ...
1:~~Also at Vienna University of Technology, Vienna, Austria\\
% ...

\end{sloppypar}
\end{document}